\documentclass{ws-procs975x65}

\begin{document}

\title{PSR J1141$-$6545: a powerful laboratory of GR and tensor-scalar
  theories of gravity}

\author{J. P. W. Verbiest$^*$}

\address{Department of Physics, West Virginia University,\\
  PO Box 6315 Morgantown, WV 26506-6315, USA\\
  $^*$E-mail: Joris.Verbiest@mail.wvu.edu}

\author{N. D. R. Bhat and M. Bailes}

\address{Centre for Astrophysics and Supercomputing, Swinburne
  University of Technology,\\
  Mail H39, PO Box 218, VIC 3122, Australia}

\begin{abstract}
  Pulsars in close binary systems have provided some of the most
  stringent tests of strong-field gravity to date. The
  pulsar--white-dwarf binary system J1141$-$6545 is specifically
  interesting due to its gravitational asymmetry, which makes it one
  of the most powerful probes of tensor-scalar theories of gravity. We
  give an overview of current gravitational tests provided by the
  J1141$-$6545 binary system and comment on how anomalous
  accelerations, geodetic precession and timing instabilities may be
  prevented from limiting future tests of gravity to come from this
  system.
\end{abstract}

\keywords{pulsars; tensor-scalar theories}

\bodymatter

\section{The J1141$-$6545 Binary System}
PSR J1141$-$6545 is a common neutron star (pulse period $P \approx
394\,$ms; spindown $\dot{P} \approx 4 \times 10^{-15}$) in a tight and
slightly eccentric orbit around a heavy white-dwarf
companion\cite{klm+00a,ts00a} (orbital period $P_{\rm b} \approx
4.7\,$hr; $e \approx 0.17$). Its orbit is close to edge-on\cite{obv02}
($i= 76^{\circ}$) and its spatial velocity is expected to be
high\cite{ts00a,obv02} ($V \approx 115\,$km/s).

\section{Overview of Gravitational Tests}
Because of the high companion mass, short orbital period and non-zero
eccentricity of the system, three relativistic effects have been
readily observed in the PSR J1141$-$6545
system\cite{bokh03,bbv08}. These three effects are the periastron
advance\cite{bbv08} $\dot{\omega} = 5.3096\pm0.0004\,^{\circ}{\rm
  yr}^{-1}$, the gravitational redshift\cite{bbv08} $\gamma = (7.73
\pm 0.11) \times 10^{-4}\,{\rm ms}$ and the orbital decay caused by
gravitational wave emission\cite{bbv08} $\dot{P}_{\rm b} = (-4.03 \pm
0.25) \times 10^{-13}$.  All three of these effects only depend on
Keplerian parameters that can be measured independently and on the
masses of the pulsar and the white-dwarf companion ($M_{\rm PSR}$ and
$M_{\rm c}$ respectively). This implies that the two most precisely measured
effects can be used to uniquely define the system by requiring $M_{\rm
  PSR} = 1.27 \pm 0.01\,{\rm M}_{\odot}$ and $M_{\rm c} = 1.02 \pm
0.01\,{\rm M}_{\odot}$. Using these values to predict $\dot{P}_{\rm
  b}^{\rm GR}$ and subsequent comparison with the measured value,
provides a test of general relativity (GR). Since GR requires the
inclination angle derived from the Shapiro delay ``shape'' parameter
to equate to that derived from the component masses and orbital
period, the orbital inclination angle derived from scintillation
studies\cite{obv02} can also be used to test GR. As described by Bhat
et al.\cite{bbv08}, GR passes both these tests without problem.

The J1141$-$6545 system is furthermore particularly powerful in
constraining tensor-scalar theories of gravity since these theories
predict significant dipolar gravitational wave emission because of the
different self-gravity of the pulsar and companion
star\cite{esp05hack}. Specifically, in the regime of strong quadrature
coupling ($\beta_0\gg 0$), timing of PSR J1141$-$6545 currently places
the strongest bound on the linear coupling constant: $\alpha_0^2 < 3.4
\times 10^{-6}$. At smaller values of $\beta_0$, its current bound is
only about a factor of three less constraining than the bound placed
by laser ranging to the Cassini spacecraft\cite{bit03}. Since the
parameters derived from timing become progressively more precise with
a longer timing baseline, the limits from PSR J1141$-$6545 are
expected to improve on the Cassini values by the middle of this
decade\cite{bbv08}. In the following section we will comment on the
effects that may constrain these efforts.

\section{Challenges in the J1141$-$6545 System}
There are three effects that could pose serious constraints on future
tests of gravity derived from timing PSR J1141$-$6545. These are
anomalous accelerations of the system, geodetic precession and
glitches, as detailed below.

\subsection{Anomalous Accelerations of the System}
Any apparent acceleration of the binary system will cause
periodicities to change as a function of time, and will hence affect
the measured orbital period derivative as well. Specifically, two
contaminating factors may prove important. First the Galactic
acceleration, both perpendicular to the Galactic plane (caused by the
Galactic gravitational potential) and within the plane (caused by
differential Galactic rotation). This effect mainly depend on the
distance of the pulsar, which has been determined\cite{obv02a} to be
larger than 3.7\,kpc. Based on that distance limit, the combined
Galactic contribution to $\dot{P}_{\rm b}$ is expected to be at most
$-5 \times 10^{-15}$. The other contaminant is the Shklovskii
effect\cite{shk70}, which depends on both the transverse velocity
$V_{\rm T}$ and the distance $D$: $\dot{P}_{\rm b}^{\rm Shk} =
\frac{V_{\rm T}^2 P_{\rm b}}{D c}$. Assuming a distance of 3.7\,kpc
and a transverse velocity\cite{obv02} of 115\,km/s, this effect is
expected to be at most of the order of $7 \times 10^{-15}$. When
compared to the current measurement precision on the orbital period
derivative: $\dot{P}_{\rm b} = -4.03 \pm 0.25 \times 10^{-13}$, it is
clear that these contaminations are still well within the precision of
our measurement and have therefore been inconsequential so
far. Accurate determination of these effects will be required,
however, to correct $\dot{P}_{\rm b}$ at the 2\% level, a precision that
should be reached by the middle of this decade. In order to enable such
a correction, VLBI observations have been proposed to place a stronger
limit on the distance and attempt an initial measurement of the proper
motion.

\subsection{Geodetic Precession}
PSR J1141$-$6545 is known to exhibit geodetic
precession\cite{hbo05,mks+10}. This effect causes changes in pulse
shape which in turn affect the timing since it biases the
cross-correlation of the pulse with a standard template. Extensive
modelling of the pulse shape and its evolution\cite{mks+10} provides a
potential correction for this effect, because the time-evolving model
of the pulse profile could be used as the basis to time the
observations against. If successful, these efforts could severely
decrease the effect geodetic precession has on the timing results.

\subsection{Glitch and Timing Noise}
Between 19 May and 15 July 2007 PSR J1141$-$6545 experienced a sudden
spin-up otherwise known as a glitch\cite{mks+10}. Since pulsar
glitches are poorly understood and because there does not exist a
general and complete model of the timing effects of pulsar
glitches\cite{mpw08}, ongoing monitoring is required to accurately
correct the glitch and its relaxation process. While this does
decrease the timing precision temporarily as the glitch model is
improved towards its final solution, it is not expected to have
lasting effect on the timing solution. The presence of this glitch
does, however, suggest that earlier timing irregularities may also
have been caused by an unmodelled glitch that occured near the time
when the pulsar was first discovered. Further investigations into this
possibility may retroactively improve the precision of the timing
solution for PSR J1141$-$6545.

\section*{Acknowledgements}
JPWV thanks M. Kramer and D. Manchester for discussions of their
recent work on PSR J1141$-$6545 and A. Melatos for interesting
discussions on pulsar glitches. JPWV also acknowledges support from a
WVEPSCoR research challenge grant held by the WVU Center for
Astrophysics.

\bibliographystyle{ws-procs975x65}

\end{document}